\magnification=1200
\overfullrule=0pt

\hsize=125mm
\vsize=180mm
\hoffset=3mm
\voffset=12mm

\baselineskip=14pt

\font\tenbb=msym10
\font\sevenbb=msym7
\font\fivebb=msym5
\newfam\bbfam
\textfont\bbfam=\tenbb \scriptfont\bbfam=\sevenbb
\scriptscriptfont\bbfam=\fivebb
\def\bb{\fam\bbfam}

\def\Cb{{\bb C}}
\def\Hb{{\bb H}}
\def\Nb{{\bb N}}
\def\Qb{{\bb Q}}
\def\Rb{{\bb R}}
\def\Tb{{\bb T}}
\def\Zb{{\bb Z}}

\def\Ac{{\cal A}}
\def\Bc{{\cal B}}
\def\Ec{{\cal E}}
\def\Hc{{\cal H}}
\def\Lc{{\cal L}}
\def\Sc{{\cal S}}
\def\Uc{{\cal U}}

\def\Aut{\mathop{\rm Aut}\nolimits}
\def\Diff{\mathop{\rm Diff}\nolimits}
\def\Dom{\mathop{\rm Dom}\nolimits}
\def\End{\mathop{\rm End}\nolimits}
\def\Hom{\mathop{\rm Hom}\nolimits}
\def\id{\mathop{\rm id}\nolimits}
\def\Ind{\mathop{\rm Ind}\nolimits}
\def\Inf{\mathop{\rm Inf}\nolimits}
\def\Int{\mathop{\rm Int}\nolimits}
\def\Proj{\mathop{\rm Proj}\nolimits}
\def\Rank{\mathop{\rm Rank}\nolimits}
\def\Sign{\mathop{\rm Sign}\nolimits}
\def\Spin{\mathop{\rm Spin}\nolimits}
\def\Sup{\mathop{\rm Sup}\nolimits}

\def\a{\alpha}
\def\b{\beta}
\def\d{\delta}
\def\fl{\forall}
\def\g{\gamma}
\def\G{\Gamma}
\def\ify{\infty}
\def\lb{\lambda}
\def\L{\Lambda}
\def\lgl{\langle}
\def\longra{\longrightarrow}
\def\mint{\int \!\!\!\!\! -}
\def\nb{\nabla}
\def\om{\omega}
\def\Om{\Omega}
\def\op{\oplus}
\def\ot{\otimes}
\def\ov{\overline}
\def\part{\partial}
\def\ra{\rightarrow}
\def\rgl{\rangle}
\def\sbs{\subset}
\def\s{\sigma}
\def\Si{\Sigma}
\def\t{\theta}
\def\ts{\times}
\def\ve{\varepsilon}
\def\vp{\varphi}
\def\wdg{\wedge}
\def\wt{\widetilde}
\def\z{\zeta}

\catcode`\@=11
\def\displaylinesno #1{\displ@y\halign{
\hbox to\displaywidth{$\@lign\hfil\displaystyle##\hfil$}&
\llap{$##$}\crcr#1\crcr}}

\def\ldisplaylinesno #1{\displ@y\halign{
\hbox to\displaywidth{$\@lign\hfil\displaystyle##\hfil$}&
\kern-\displaywidth\rlap{$##$}
\tabskip\displaywidth\crcr#1\crcr}}
\catcode`\@=12

\def\buildrel#1\over#2{\mathrel{
\mathop{\kern 0pt#2}\limits^{#1}}}

\def\build#1_#2^#3{\mathrel{
\mathop{\kern 0pt#1}\limits_{#2}^{#3}}}

\def\tvi{\vrule height 12pt depth 5pt width 0pt}

\def\tv{\tvi\vrule}

\def\cc#1{\hfill\kern .7em#1\kern .7em\hfill}

\def\up#1{\raise 1ex\hbox{\sevenrm#1}}

\rightline{hep-th/9603053}

\vglue 1cm

\centerline{\bf Gravity coupled with matter and the}

\centerline{\bf foundation of non commutative geometry}

\bigskip

\centerline{Alain CONNES}

\vglue 0.3cm

\centerline{I.H.E.S., 91440 Bures-sur-Yvette, France}
\vglue 1cm
\noindent {\parindent=1cm\narrower
{\bf Abstract.} We first exhibit in the commutative case
the simple algebraic relations between the algebra of
functions on a manifold and its infinitesimal length
element $ds$. Its unitary representations correspond to
Riemannian metrics and Spin structure while $ds$ is the
Dirac propagator $ds = \ts \!\!$---$\!\! \ts = D^{-1}$
where $D$ is the Dirac operator.

\noindent We extend these simple relations to the non
commutative case using Tomita's involution $J$. We then
write a spectral action, the trace of a function of the
length element in Planck units, which when applied to the
non commutative geometry of the Standard Model will be
shown (in a joint work with Ali Chamseddine) to give the
SM Lagrangian coupled to gravity. The internal
fluctuations of the non commutative geometry are trivial
in the commutative case but yield the full bosonic sector
of SM with all correct quantum numbers in the slightly
non commutative case. The group of local gauge
transformations appears spontaneously as a normal subgroup
of the diffeomorphism group. \par}

\vglue 2cm

Riemann's concept of a geometric space is based on the
notion of a manifold $M$ whose points $x\in M$ are
locally labelled by a finite number of real coordinates
$x^{\mu} \in \Rb$. The metric data is given by the
infinitesimal length element,
$$
ds^2 = g_{\mu \nu} \, dx^{\mu} \, dx^{\nu} \leqno (1)
$$
which allows to measure distances between two points
$x,y$ as the infimum of the length of arcs $x(t)$ from
$x$ to $y$,
$$
d(x,y) = \Inf \int_x^y ds \, . \leqno (2)
$$
In this paper we shall build our notion of geometry, in a
very similar but somehow dual manner, on the pair $(\Ac
,ds)$ of the algebra $\Ac$ of coordinates and the
infinite\-simal length element $ds$. For the start we
only consider $ds$ as a symbol, which together with $\Ac$
generates an algebra $(\Ac ,ds)$. The length element $ds$
does not commute with the coordinates, i.e. with the
functions $f$ on our space, $f\in \Ac$. But it does
satisfy non trivial relations. Thus in the simplest case
where $\Ac$ is commutative we shall have, 
$$
\left[[f,ds^{-1}] ,g\right] = 0 \qquad \fl \, f,g \in \Ac
\, . \leqno (3) 
$$
The only other relation between $ds$ and $\Ac$ is of
degree $n$ in $ds^{-1} =D$ and expresses the homological
nature of the volume form (see axiom 4 below). When $\Ac$
is commutative it has a spectrum, namely the space of
algebra preserving maps from $\Ac$ to the complex numbers,
$$ 
\ldisplaylinesno{ \chi : \Ac \ra \Cb \ ; \ \chi (a+b) =
\chi (a) + \chi (b) \ , \ \chi (ab) = \chi (a) \, \chi (b)
\quad \fl \, a,b \in \Ac \ , &(4) \cr
\chi (\lb a) = \lb \, \chi (a) \quad \fl \, \lb \in \Cb \
, \ \fl \, a \in \Ac \, . \cr
}
$$
For instance, when $\Ac$ is the algebra of functions on a
space $M$ the space of such maps, called {\it characters}
of $\Ac$ identifies with $M$. To each point $x\in M$
corresponds the character $\chi_x$,
$$
\chi_x (f) = f(x) \qquad \fl \, f\in \Ac \, . \leqno (5)
$$
While relations such as (3) between the algebra $\Ac$ and
the length element $ds$ hold at the universal level, a
specific geometry will be specified as a {\it unitary
representation} of the algebra generated by $\Ac$ and
$ds$. In general we shall deal with complex valued
functions $f$ so that $\Ac$ will be endowed with an
involution,
$$
f \ra f^* \leqno (6)
$$
which is just complex conjugation of functions in the
usual case,
$$
f^* (x) = \ov{f(x)} \qquad \fl \, x\in M \, . \leqno (7)
$$
The length element $ds$ will be selfadjoint,
$$
ds^* = ds \leqno (8)
$$
so that $(ds)^2$ will be automatically positive.

\smallskip

\noindent The unitarity of the representation just means
that the operator $\pi (a^*)$ correspon\-ding to $a^*$ is
the adjoint of the operator $\pi (a)$,
$$
\pi (a^*) = \pi (a)^* \qquad \fl \, a \in (\Ac ,ds) \, .
\leqno (9)
$$
Given a unitary representation $\pi$ of $(\Ac ,ds)$ we
measure the distance between two points $x,y$ of our
space by,
$$
d(x,y) = \Sup \left\{ \vert f(x) - f(y)\vert \ ; \ f\in
\Ac \ , \ \Vert [f,ds^{-1}]\Vert \leq 1 \right\} \leqno
(10)
$$
where we dropped the letter $\pi$ but where the
representation $\pi$ has been used in a crucial way to
define the norm of $[f,ds^{-1}]$.

\smallskip

\noindent Before we proceed let us first explain what
representation of $(\Ac ,ds)$ corresponds to an ordinary
Riemannian geometry, and check that (10) gives us back
exactly the Riemannian geodesic distance of formula (2).
The algebra $\Ac$ is the algebra of smooth complex valued
functions on $M$, $\Ac = C^{\ify} (M)$ and we choose to
represent $ds$ as the propagator for fermions which
physicists write in a suggestive ways as $\ts
\!\!$---$\!\! \ts$. This means that we represent $\Ac$ in
the Hilbert space $L^2 (M,S)$ of square integrable
sections of the spinor bundle on $M$ by the following
formula,  $$
(f \, \xi)(x) = f(x) \, \xi (x) \qquad \fl \, f \in \Ac =
C^{\ify} (M) \ , \ \fl \, \xi \in \Hc = L^2 (M,S) \leqno
(11)
$$
and we represent $ds$ by the formula,
$$
ds = D^{-1} \ , \ D={1\over \sqrt{-1}} \, \g^{\mu} \,
\nb_{\mu} \leqno (12)
$$
where $D$ is the Dirac operator. We ignore the ambiguity
of (12) on the kernel of $D$.

\smallskip

\noindent One checks immediately that the commutator
$[D,f]$, for $f\in \Ac = C^{\ify} (M)$ is the Clifford
multiplication by the gradient $\nb f$ so that its
operator norm, in $\Hc = L^2 (M,S)$, is given by
$$
\Vert [D,f]\Vert = \ \build \Sup_{x\in M}^{} \Vert \nb
f\Vert \, . \leqno (13)
$$
It then follows by integration along the path from $x$ to
$y$ that $\vert f(x) - f(y)\vert \leq$ length of path,
provided (13) is bounded by 1. Thus the equality between
(10) and (2).

\smallskip

\noindent Note that while $ds$ has the dimension of a
length, $(ds)^{-1}$ which is represented by $D$ has the
dimension of a mass. The formula (10) is dual to formula
(2). In the usual Riemannian case it gives the same
answer but being dual it does not use arcs connecting $x$
with $y$ but rather functions from $M$ to $\Cb$. As we
shall see this will allow to treat spaces with a finite
number of points on the same footing as the continuum.
Another virtue of formula (10) is that it will continue
to make sense when the algebra $\Ac$ is no longer
commutative.

\smallskip

\noindent In this paper we shall write down the axioms of
geometry as the presentation of the algebraic relations
between $\Ac$ and $ds$ and the representation of those
relations in Hilbert space.

\smallskip

\noindent In order to compare different geometries, i.e.
different representations of the algebra $(\Ac ,ds)$
generated by $\Ac$ and $ds$, we shall use the following
action functional,
$$
\hbox{Trace} \, \left( \vp \left( {ds \over \ell_p}
\right) \right) \leqno (14)
$$
where $\ell_p$ is the Planck length and $\vp$ is a
suitable cutoff function which will cut off all
eigenvalues of $ds$ larger than $\ell_p$.

\smallskip

\noindent We shall show in [CC] that for a suitable
choice of the algebra $\Ac$, the above action will give
Einstein gravity coupled with the Lagrangian of the
standard $U(1) \ts SU(2) \ts SU(3)$ model of Glashow
Weinberg Salam. The algebra will not be $C^{\ify} (M)$
with $M$ a (compact) 4-manifold but a non commutative
refinement of it which has to do with the quantum group
refinement of the Spin covering of $SO(4)$,
$$
1 \ra \Zb /2 \ra \Spin (4) \ra SO(4) \ra 1 \, .\leqno (15)
$$
Amazingly, in this description the group of gauge
transformations of the matter fields arises spontaneously
as a normal subgroup of the generalized diffeomorphism
group $\Aut (\Ac)$.

\smallskip

\noindent It is the {\it non commutativity} of the algebra
$\Ac$ which gives for free the group of gauge
transformations of matter fields as a (normal) subgroup of
the group of diffeomorphisms. Indeed, when $\Ac = C^{\ify}
(M)$ is the commutative algebra of smooth functions on $M$
one easily checks that the following defines a one to one
correspondence between {\it diffeomorphisms} $\vp$ of $M$
and the {\it automorphisms} $\a \in \Aut (\Ac)$ of the
algebra $\Ac$ (preserving the $*$),
$$
\a (f) \, (x) = f(\vp^{-1} (x)) \qquad \fl \, x\in M \ , \
f\in C^{\ify} (M) \, . \leqno (16)
$$
In the non commutative case an algebra always has
automorphisms, the {\it inner automorphisms} given
by
$$
\a_u (f) = u \, f \, u^* \qquad \fl \, f \in \Ac \leqno
(17)
$$
for any element of the unitary group $\Uc$ of $\Ac$,
$$
\Uc = \{ u \in \Ac \ ; \ u \, u^* = u^* \, u =1 \} \, .
\leqno (18)
$$
The subgroup $\Int (\Ac) \sbs \Aut (\Ac)$ of inner
automorphisms is a normal subgroup and it will provide us
with our group of internal gauge transformations. It is a
happy coincidence that the two terminologies : inner
automorphisms and internal symmetries actually match. Both
groups will have to be lifted to the spinors but we shall
see that later. We shall see that the action of inner
automorphisms on the metric gives rise to {\it internal
fluctuations} of the latter which replace $D$ by $D+A+J
\, A \, J^{-1}$ (see below) and give exactly the gauge
bosons of the standard model, with its
symmetry breaking Higgs sector, when we apply it to the
above finite geometry. We had realized in [Co2] that the
fermions were naturally in the adjoint representation of
the unitary group $\Uc$ in the case of the standard
model, i.e. that the action of $\Uc$ on $\Hc$ was,
$$
\xi \ra u \, J \, u \, J^{-1} \, \xi = u \, \xi \, u^* \,
.
$$
But we had not understood the significance of the inner
automorphisms as internal diffeomorphisms of the non
commutative space. What the present paper shows is that
one should consider the internal gauge symmetries as part
of the diffeomorphism group of the non commutative
geometry and the gauge bosons as the internal
fluctuations of the metric. We should thus expect that
the action functional should be of purely gravitational
nature. The above spectral action when restricted to these
special metrics will give the interaction Lagrangian of the
bosons. The interaction with Fermions, which has the right
hypercharge assignement is obtained directly from $\lgl
\psi ,D\psi \rgl$ (with $D+A+J \, A \, J^{-1}$ instead of
$D$), and is thus also of spectral nature being invariant
under the full unitary group of operators in Hilbert
space. Thus the only distinction that remains between
matter and gravity is the distinction $\Int \Ac \ne \Aut
\Ac$ which vanishes for a number of highly non commutative
algebras.

\bigskip

\noindent {\it Quantized calculus.}

\smallskip

\noindent In order to ease the presentation of our axioms
let us recall the first few lines of the dictionary of
the quantized calculus which makes use of the formalism
of quantum mechanics to formulate a new theory of
infinitesimals. The first two lines of this dictionary
are just the traditional way of interpreting the
observables in quantum mechanics,
$$
\matrix{
\hbox{\bf Classical} &\hbox{\bf Quantum} \cr
\cr
\hbox{Complex variable} &\hbox{Operator in Hilbert space}
\cr
\hbox{Real variable} &\hbox{Selfadjoint operator,} \
T=T^* \cr
\hbox{Infinitesimal variable} &\hbox{Compact operator} \cr
\hbox{Infinitesimal of order} \ \a &\mu_n (T) =
0(n^{-\a}) \cr
\hbox{Integral} &\mint \, T=\log \ \hbox{divergence of the
trace of} \ T \cr
}
$$

\noindent We recall briefly that an operator $T$ in Hilbert
space is {\it compact} iff for any $\ve >0$ one has $\Vert
T \Vert < \ve$ except on a finite dimensional subspace of
$\Hc$. More precisely
$$
\ldisplaylinesno{
\fl \, \ve \ \hbox{there exists a finite dimensional
subspace} \ E \ \hbox{of} \ \Hc &(19) \cr 
\hbox{such that} \ \Vert T / E^{\perp} \Vert < \ve \quad
\hbox{where} \ E^{\perp} \ \hbox{is the orthogonal of} \
E \, . \cr  
}
$$
The size of a compact operator $T$ is measured by the
decreasing sequence $\mu_n$ of eigenvalues of $\vert T
\vert = \sqrt{T^* \, T}$. The order of such an
``infinitesimal'' is measured by the rate at which these
{\it characteristic values} $\mu_n (T)$ converge to $0$
when $n\ra \ify$.

\smallskip

\noindent One can show that all the intuitive rules of
calculus are valid, e.g. the order of $T_1 + T_2$ or of
$T_1 \, T_2$ are as they should be ($\leq \a_1 \vee \a_2$
and $\a_1 \, \a_2$). Moreover the trace (i.e. the sum of
the eigenvalues) is logarithmically divergent for
infinitesimals of order 1, since $\mu_n (T) = 0
\left({1\over n}\right)$. It is a quite remarkable fact
that the coefficient of the logarithmic divergency does
yield an {\it additive trace} which in essence eva\-luates
the ``classical part'' of such infinitesimals. This
trace, denoted $\mint$ vanishes on infinitesimals of
order $\a > 1$.

\smallskip

\noindent The only rule of the naive calculus of
infinitesimals which is not fulfilled is commutativity
but this lack of commutativity is crucial to allow the
coexistence of variables with continuous range with
infinitesimals which have discrete spectrum.

\smallskip

\noindent We refer to [Co] for more details on this
calculus and its applications.

\bigskip

\noindent {\it Axioms for commutative geometry.}

\smallskip

\noindent Let us now proceed and write down the axioms
for {\it commutative} geometry. The small modifications
required for the non commutative case will only be
handled later.

\smallskip

\noindent Thus $\Hc$ is a Hilbert space, $\Ac$ an
involutive algebra represented in $\Hc$ and $D=ds^{-1}$
is a selfadjoint operator in $\Hc$.

\smallskip

\noindent We are given an integer $n$ which controls the
dimension of our space by the condition,

\smallskip

\noindent 1) {\it $ds = D^{-1}$ is an infinitesimal of
order ${1\over n}$.}

\smallskip

\noindent The universal commutation relation (3) is
represented by

\smallskip

\noindent 2) {\it $\left[ [D,f],g\right] = 0 \qquad \fl \,
f,g \in \Ac$.}

\smallskip

\noindent We assume that the simple commutation with
$\vert D \vert , \d (T) = [\vert D \vert ,T]$ will only
yield bounded operators when we start with any $f\in
\Ac$. More precisely we assume that:

\smallskip

\noindent 3) {\it (Smoothness) For any $a\in \Ac$ both $a$
and $[D,a]$ belong to the domain of $\d^m$, for any integer
$m$.}

\smallskip

\noindent This axiom 3 is the algebraic formulation of
smoothness of the coordinates.

\smallskip

\noindent The next axiom is depending upon the parity of
$n$, thus let us state it first when $n$ is even. It
yields the $\g_5$ matrix which is abstracted here as a
$\Zb /2$ grading of the Hilbert space $\Hc$ :

\smallskip

\noindent 4) {\it (Orientability) ($n$ even) There exists a
Hochschild cycle $c \in Z_n (\Ac ,\Ac)$ such that $\pi
(c) =\g$ satisfies
$$
\g = \g^* \ , \ \g^2 = 1 \ , \ \g D =-D\g \, .
$$
In the odd case ($n$ odd) one just asks that $1=\pi (c)$
for some $n$-dimensional Hochschild cycle $c\in Z_n (\Ac
,\Ac)$.}

\smallskip

\noindent We need to explain briefly what is a Hochschild
cycle. Conceptually it is the algebraic formulation of a
differential form, so that axiom 4 is really providing us
with the volume form. Concretely an $n$-dimensional
Hochschild cycle is a finite sum of elements of $\Ac \ot
\Ac \ot \ldots \ot \Ac$ (with $n+1$ times $\Ac$),
$$
c = \sum_{j}^{} a_j^0 \ot a_j^1 \ot a_j^2 \ldots \ot a_j^n
$$
such that the contraction $bc=0$ where by definition $b$
is linear and satisfies
$$
\displaylines{
b(a^0 \ot a^1 \ot \ldots \ot a^n) = a^0 a^1 \ot a^2 \ot
\ldots \ot a^n - a^0 \ot a^1 a^2 \ot \ldots \ot a^n \cr
+ \ldots + (-1)^k a^0 \ot \ldots \ot a^k a^{k+1} \ot
\ldots \ot a^n + \ldots + (-1)^n a^n a^0 \ot \ldots \ot
a^{n-1} \, . \cr
}
$$
When $\Ac$ is commutative it
is easy to construct a Hochschild cycle, it suffices to
take any $a^j$ and consider
$$
c = \sum \ve (\s) \, a^0 \ot a^{\s (1)} \ot a^{\s (2)}
\ot \ldots \ot a^{\s (n)}
$$
where the sum runs over all the permutations $\s$ of $\{
1,\ldots ,n\}$. 

\smallskip

\noindent This special construction corresponds to
the familiar differential form $a^0 da^1 \wdg da^2 \wdg
\ldots \wdg da^n$ but does not require the previous
knowledge of the tangent bundle.

\smallskip

\noindent Finally $\pi (c)$ is the representation of $c$
on $\Hc$ induced by
$$
\pi (a^0 \ot a^1 \ot \ldots \ot a^n) = a^0 [D,a^1] \ldots
[D,a^n] \, .
$$

\smallskip

\noindent To understand the meaning of axiom 4 let us
take its simplest instance, with $n=1$ and $\Ac =
C^{\ify} (S^1)$ generated by a single unitary element $U
\in \Ac$, $U \, U^* = U^* \, U =1$. Let $c = U^{-1} \ot
U$ one checks that $bc = U^{-1} \, U - U \, U^{-1} =0$
so that $c$ is a Hochschild cycle, $c\in Z^1 (\Ac ,\Ac)$.
Then the condition $\pi (c) =1$ reads:
$$
[D,U] = U \, .
$$
The reader will check at this point that this
relation alone completely describes the geometry of the
circle, by computing explicitly the distance between
points using formula (10). One finds the usual metric on
the circle with length $2\pi$.

\smallskip

\noindent The next axiom will be the axiom of {\it
finiteness},

\medskip

\noindent 5) {\it (Finiteness and absolute continuity.)
Viewed as an $\Ac$-module the space $\Hc_{\ify} = \,
\build \cap_{m}^{} \, \hbox{\rm Domain} \, D^m$ is}
finite and projective. {\it Moreover the following
equality defines a hermitian structure ( , ) on this
module,} 
$$
\lgl a \, \xi ,\eta \rgl = f \, a (\xi ,\eta) \, ds^n
\qquad \fl \, a \in \Ac \ , \ \fl \, \xi ,\eta \in
\Hc_{\ify} \, .
$$

\medskip

\noindent We recall from [Co] that a hermitian structure on
a finite projective $\Ac$-module is given by an
$\Ac$-valued inner product. The prototype of such a
module with inner product is the following, one lets
$e\in \Proj M_q (\Ac)$ be a selfadjoint idempotent,
$$
e=e^* \ , \ e^2 =e
$$
and one lets $\Ec$ be the left module $\Ac^n \, e = \{
(\xi_j)_{j=1,\ldots ,n} \, ; \, \xi_j \in \Ac \, , \,
\xi_j \, e_{jk} = \xi_k \}$ where the action of $\Ac$ is
given by left multiplication,
$$
(a \, \xi)_j = a \, \xi_j \qquad \fl \, j=1,\ldots ,n \,
.
$$

\noindent The $\Ac$-valued inner product is then given by:
$$
(\xi ,\eta) = \Si \, \xi_i \, \eta_i^* \, .
$$
It follows from axiom 4) and from a general theorem ([Co])
that the operators $a \, ds^n$, $a \in \Ac$ are {\it
measurable} (cf. [Co]) so that the coefficient $\mint
\, a \, ds^n$ of the logarithmic divergence of their trace
is unambiguously defined.

\smallskip

\noindent It follows from axiom 5) that the algebra $\Ac$
is uniquely determined inside its weak closure $\Ac''$
(which is also the bicommutant of $\Ac$ in $\Hc$) by the
equality
$$
\Ac = \{ T\in \Ac'' \ ; \ T\in \, \build \cap_{m>0}^{} \,
\Dom \d^m \} \, .
$$
This shows that the whole geometric data $(\Ac ,\Hc ,D)$
is in fact uniquely determined by the triple $(\Ac'' ,\Hc
,D)$ where $\Ac''$ is a commutative von Neumann algebra.

\smallskip

\noindent This also shows that $\Ac$ is a pre $C^*$
algebra, i.e. that it is stable under the $C^{\ify}$
functional calculus in the $C^*$ algebra norm closure of
$\Ac$, $A=\ov{\Ac}$. Since we assumed that $\Ac$ was
commutative, so is $A$ and by Gelfand's theorem $A =
C(X)$ is the algebra of continuous complex valued
functions on $X = \hbox{Spectrum} \, (A)$. We note
finally that characters $\chi$ of $\Ac$ extend
automatically to $A$ by continuity so that
$$
\hbox{Spectrum} \, A = \hbox{Spectrum} \, \Ac \, .
$$
We let $K_i (\Ac)$, $i=0,1$ be the $K$-groups of $\Ac$
(or equivalently of $A$ or of $X$). Thus $K_0 (\Ac)$
classifies finite projective modules over $\Ac$ (or
equivalently vector bundles over $X$). Similarly $K_1
(\Ac)$ is the group of connected components of $GL_{\ify}
(\Ac)$, i.e. $K_1 (\Ac) = \pi_0 \, GL_{\ify} (\Ac)$,
while by the Bott periodicity theorem one has,
$$
\pi_n (GL_{\ify} (\Ac)) \cong K_{n+1} (\Ac)
$$
where $n+1$ only matters modulo 2.

\smallskip

\noindent One easily defines the index pairing of the
operator $D$ with $K_n (\Ac)$ where again $n$ only
matters modulo 2. In the even case one uses $\g$ to
decompose $D$ as $D=D^+ + D^-$ where $D^+ = (1-p) \, Dp$,
$p={1+\g \over 2}$. Then for any selfadjoint idempotent
$e\in \Ac$ the operator $e \, D^+ \, e$ is a Fredholm
operator from the subspace $e \, p \, \Hc$ of $\Hc$ to
the subspace $e(1-p) \Hc$. This extends immediately to
projections $e\in M_q (\Ac)$ for some integer $q$, and
gives an additive map from $K_0 (\Ac)$ to $\Zb$ denoted
$\lgl\Ind D, e\rgl$. A similar discussion applies in the
odd case ([]).

\smallskip

\noindent Since $\Ac$ is commutative we have the
diagonal map,
$$
\Ac \ot \Ac \build \longra_{}^{m} \Ac \ , \ m(x\ot y) =
xy \qquad \fl \, x,y \in \Ac
$$
which yields a corresponding map $m_* : K_* (\Ac) \ts
K_* (\Ac) \ra K_* (\Ac)$. Composing this map with $\Ind
D$ we obtain the intersection form,
$$
K_* (\Ac) \ts K_* (\Ac) \ra \Zb \, ,
$$
$$
(e,f) \ra \lgl \Ind D \ , \ m_* (e\ot f)\rgl \, .
$$
It clearly only depends upon the stable homotopy class of
the representation $\pi$ and thus gives a very rough
information on $\pi$. We shall assume, (Poincar\'e
duality)

\medskip

\noindent 6) {\it The intersection form $K_* (\Ac) \ts
K_* (\Ac) \ra \Zb$ is invertible.}

\medskip

\noindent If one wants to take in account the possible
presence of torsion in the $K$-groups one should
formulate Poincar\'e duality as the isomorphism,
$$
K_* (\Ac) \build \longra_{}^{\cap \mu} K^* (\Ac)
$$
given by the Kasparov intersection product with the class
$\mu$ of the Fredholm module $(\Hc ,D,\g)$ over $\Ac \ot
\Ac$. We refer to [Co] [C] for the detailed formulation,
but 6) will suffice for theorem 1 below.

\smallskip

\noindent Thanks to the work of D. Sullivan [S] one knows
that the above Poincar\'e duality isomorphism in $K$
theory is (in the simply connected case and ignoring
2-torsion) a characterization of the homotopy type of
spaces which possess a structure of smooth manifold. This
requires the use of real $K$ theory instead of the above
complex $K$ theory and our last axiom will be precisely
the existence of such a real structure on our cycle.

\medskip

\noindent 7) {\bf Reality.} {\it There exists an
antilinear isometry $J: \Hc \ra \Hc$ such that $J \, a \,
J^{-1} = a^* \qquad \fl \, a\in \Ac$ and $J^2 =\ve$, $J
\, D = \ve' \, D \, J$ and $J \, \g = \ve'' \, \g \, J$
where $\ve ,\ve' ,\ve'' \in \{ -1,+1\}$ are given by the
following table from the value of $n$ modulo 8,}

$$
\vbox{
\offinterlineskip
\halign{
\tv#&#   &\tv#&#  &\tv#&#  &\tv#&#  &\tv#&#  
&\tv#&#  &\tv#&#  &\tv#&# &\tv#&# &\tv#&# 
&\tv#\cr
\noalign{\hrule}
\cc{$n=$} &&\cc{$0$} &&\cc{$1$} &&\cc{$2$} &&\cc{$3$} 
&&\cc{$4$} &&\cc{$5$} &&\cc{$6$} &&\cc{$7$} &&\cr 
\cc{$\ve$} &&\cc{$1$} &&\cc{$1$} &&\cc{$-1$} 
&&\cc{$-1$} &&\cc{$-1$} &&\cc{$-1$} &&\cc{$1$} 
&&\cc{$1$} &&\cr
\cc{$\ve'$} &&\cc{$1$} &&\cc{$-1$} &&\cc{$1$} &&\cc{$1$} 
&&\cc{$1$} &&\cc{$-1$} &&\cc{$1$} &&\cc{$1$} &&\cr
\cc{$\ve''$} &&\cc{$1$} && &&\cc{$-1$} 
&& &&\cc{$1$} && &&\cc{$-1$} && &&\cr
\noalign{\hrule} }}
$$

\medskip

\noindent We can now state our first result,

\medskip

\noindent {\bf Theorem.} {\it Let $\Ac = C^{\ify} (M)$
where $M$ is a smooth compact manifold of dimension $n$.}
a) {\it Let $\pi$ be a unitary representation of $(\Ac
,ds)$ satisfying the above seven axioms, then there exists
a unique Riemannian metric $g$ on $M$ such that the
geodesic distance between any two points $x,y \in M$ is
given by} 
$$
d(x,y) = \Sup \, \{ \vert a(x) - a(y)\vert \ ; \ a\in \Ac \
, \ \Vert [D,a]\Vert \leq 1 \} \, .
$$

\noindent b) {\it The metric $g=g(\pi)$ only depends upon
the unitary equivalence class of $\pi$ and the fibers of
the map : $\pi \ra g(\pi)$ from unitary equivalence classes
to metrics form a finite collection of affine space
$\Ac_{\s}$ parametrized by the Spin structures $\s$ on
$M$.}

\smallskip

\noindent c) {\it The action functional $\mint \, ds^{n-2}$
is a positive quadratic form on each $\Ac_{\s}$ with a
unique minimum $\pi_{\s}$.}

\smallskip

\noindent d) {\it $\pi_{\s}$ is the representation of
$(\Ac ,ds)$ in $L^2 (M,S_{\s})$ given by multiplication
operators and the Dirac operator of the Levi Civita Spin
connection.}

\smallskip

\noindent e) {\it The value of $\mint \, ds^{n-2}$ on
$\pi_{\s}$ is given by the Einstein Hilbert action,}
$$
-c_n \int R \, \sqrt{g} \, d^n x \quad , \quad c_n =
(n-2)/12 \ts (4\pi)^{-n/2} \, \G \left( {n\over 2} +1
\right)^{-1} \, 2^{[n/2]} \, . 
$$

\medskip

\noindent Let us make a few remarks about this theorem.

\smallskip

\item{1)} Note first that none of the axioms uses the
fact that $\Ac$ is the algebra of smooth functions on a
manifold. In fact one should deduce from the axioms that
the spectrum $X$ of $\Ac$ is a smooth manifold and that
the map $X \ra \Rb^N$, given by the finite collection
$a_j^i$ of elements of $\Ac$ involved in the Hochschild
cycle $c$ of axiom 4, is actually a smooth embedding of
$X$ as a submanifold of $\Rb^N$. To prove this one should
use [Co], proposition 15, p.312.

\smallskip

\item{2)} The two axioms 2 and 3 should be considered as
the presentation of the algebra $(\Ac ,ds)$. Thus this
presentation involves an explicit Hochschild $n$-cycle
$c\in Z_n (\Ac ,\Ac)$ and for $n$ odd gives the relation,
$$
\Si \, a_0^j [D,a_1^j] \ldots [D,a_n^j] =1 \, .
$$
In the even case the relations are $D \, \g = -\g \, D$
and $\g^2 =1$, $\g = \g^*$, with
$$
\g = \Si \, a_0^j [D,a_1^j] \ldots [D,a_u^j] \, .
$$
It is thus natural to fix $c$ as part of the data. The
differential form on $M= \hbox{Spectrum} \, \Ac$
associated to the Hochschild $n$-cycle $c$ is equal to the
volume form of the metric $g$. Thus fixing $c$ determines
the volume form of the metric.

\smallskip

\item{3)} The sign in front of the Einstein Hilbert
action given in (a) is the correct one for the Euclidean
formulation of gravity and for instance in the $n = 4$
dimensional case the Einstein Hilbert action becomes the
{\it area} 
$$
\ell_p^{-2} \, {\textstyle \mint} \, ds^2
$$ 
of our space, in units of Planck length $\ell_p$. Thus the
only negative sign in the second derivative of $\mint \,
ds^2$ around flat space comes from the Weyl factor which
is determined by the choice of $c$. We refer to [K] and
[KW] for the detailed calculation.

\smallskip

\item{4)} When $M$ is a Spin manifold the map $\pi \ra
g(\pi)$ is surjective and if we fix $c$ it surjects to
the metrics $g$ with fixed volume form. (This amounts to
checking that all axioms are fulfilled.)

\smallskip

\item{5)} If we drop axiom 7 there is a completely
similar result as theorem 1 where Spin is replaced by
$\hbox{Spin}^c$ (cf. [ML]) and where uniqueness is lost in
c) and the minimum of the action $f \, ds^{n-2}$ is
reached on a linear subspace of $\Ac_{\s}$ with $\s$ a
fixed $\hbox{Spin}^c$ structure. The elements of this
subspace correspond exactly to the $U(1)$ gauge potentials
involved in the $\hbox{Spin}^c$ Dirac operator and d) e)
continue to hold.

\smallskip

\item{6)} A commutative geometry $(\Ac ,\Hc ,D)$ is
connected iff the only operators commuting with $\Ac$ and
$D$ are the scalars, i.e. iff the representation $\pi$ of
$(\Ac ,ds)$ in $\Hc$ is irreducible.

\bigskip

\noindent {\it The axioms for non commutative geometry.}

\smallskip

We are now ready to proceed to the non commutative case,
i.e. where we no longer assume that the algebra $\Ac$ is
commutative. Among the axioms we wrote 1) 3) 5) will
remain unchanged while the others will be slightly
modified. One of the most significant results of the
theory of operator algebras is Tomita's theorem [Ta]
which asserts that for any weakly closed $*$ algebra of
operators $M$ in Hilbert space $\Hc$ which admits a
cyclic and separating vector, there exists a canonical
antilinear isometric involution $J$ from $\Hc$ to $\Hc$
such that
$$
J \, M \, J^{-1} = M'
$$
where $M' = \{ T \ ; \ T \, a = a \, T \quad \fl \, a
\in M\}$ is the commutant of $M$. It follows then that
$M$ is antiisomorphic to its commutant, the
antiisomorphism being given by the $\Cb$-linear map,
$$
a \in M \ra J \, a^* \, J^{-1} \in M' \, .
$$
Now axiom 7 already involves an antilinear isometry $J$
in $\Hc$, which in the usual \break geometric case is the
charge conjugation. Since we assumed that $\Ac$ was
commutative the equality $J \, a^* \, J^{-1} =a$ of axiom
7 is compatible with Tomita's antiisomorphism. In general
we shall replace the requirement $J \, a^* \, J^{-1} =a
\quad \fl \, a \in \Ac$ of axiom 7 by the following,

\medskip

\noindent $7'$) One has $[a,b^0] = 0 \quad \fl \, a,b \in
\Ac$ where $b^0 = J \, b^* \, J^{-1}$ (and $[a,b^0]$ is
the commutator $ab^0 - b^0 a$).

\smallskip

\noindent Otherwise we leave 7) unchanged. This
immediately turns the Hilbert space $\Hc$ into a module
over the algebra $\Ac \ot \Ac^0$ which is the tensor
product of $\Ac$ by its opposite algebra $\Ac^0$. One
lets,
$$
(a\ot b^0) \, \xi = a \, J \, b^* \, J^{-1} \, \xi \qquad
\fl \, a,b \in \Ac \, .
$$
(One can use equivalently the terminology of bimodules or
correspondences (cf. [Co]).)

\smallskip

\noindent Thus axiom ($7'$) now gives a $KR$-homology class
for the algebra $\Ac \ot \Ac^0$ endowed with the
antilinear automorphism,
$$
\tau (x\ot y^0) = y^* \ot x^{*0} \qquad \fl \, x,y \in \Ac
$$
and we do not need the diagonal map $m:\Ac \ot \Ac \ra
\Ac$ (which is an algebra homomorphism only in the
commutative case) to formulate Poincar\'e duality,

\medskip

\noindent ($6'$) {\it The cup product by $\mu \in KR^n (\Ac
\ot \Ac^0)$ is an isomorphism}
$$
K_* (\Ac) \build \longra_{}^{\cap \mu} K^* (\Ac) \, .
$$
We also note that the intersection form on $K_* (\Ac)$
continues to be well defined. Given $e,f \in K_* (\Ac)$
one considers $e\ot f^0$ as an element of $K_* (\Ac \ot
\Ac^0)$ and evaluates
$$
\lgl e,f \rgl = \lgl \hbox{Index} \ D \ , \ e \ot f^0
\rgl \, . 
$$
From the table of commutation of axiom 7 one gets for
instance that this intersection form is symplectic for
$n=2$ or 6 and quadratic for $n=0$ or 4 modulo 8 as in
the usual case.

\smallskip

\noindent We shall modify axiom 2 in the following way,
$$
[[D,a], b^0] = 0 \qquad  \fl \, a,b \in \Ac \, . \leqno
(2')
$$
Note that by ($7'$) $a$ and $b^0$ commute so that the
formulation of ($2'$) is symmetric, i.e. it is equivalent
to $$
[[D,b^0], a] = 0 \qquad \fl \, a,b \in \Ac \, .
$$
Finally we shall slightly modify (4). Of course
Hochschild homology continues to make sense in the non
commutative case.

\medskip

\noindent ($4'$) {\it There exists a Hochschild cycle $c\in
Z_n (\Ac ,\Ac \ot \Ac^0)$ such that $\g = \pi (c)$
satisfies $\g =\g^*$, $\g^2 =1$, $\g \, a = a \, \g \quad
\fl \, a\in \Ac$, $\g \, D = -D \, \g$. (In the odd case we
simply require $\pi (c) = 1$.)}

\smallskip

\noindent We view $\Ac \ot \Ac^0$ as a bimodule over
$\Ac$ by restricting to the subalgebra $\Ac \ot 1 \sbs
\Ac \ot \Ac^0$ the natural structure of $\Ac \ot \Ac^0$
bimodules on $\Ac \ot \Ac^0$. This gives
$$
a (b \ot c^0) \, d = a \, b \, d \ot c^0 \qquad \fl \,
a,b,c,d \in \Ac \, .
$$
Note that the Hochschild homology makes sense with
coefficients in any bimodule. Since we have a
representation of $\Ac \ot \Ac^0$ in $\Hc$, $\pi (c)$
continues to make sense.

\smallskip

\noindent The axioms (3) and (5) are unchanged in the non
commutative case, and the proof of the measurability of
the operators $a \, ds^n$ for any $a\in \Ac$ continues to
hold in that generality.

\smallskip

\noindent Thus a {\it non commutative geometry} of
dimension $n$ is given by a triple $(\Ac ,\Hc ,D)$ with
real structure $J$ satisfying (1) ($2'$) (3) ($4'$) (5)
($6'$) ($7'$).

\bigskip

\noindent {\it Examples and internal diffeomorphisms.}

\smallskip

\noindent Let us give some simple examples. Let us first
assume that $\Ac$ is the finite dimensional commutative
algebra $\Ac = \Cb^N$. Even though $\Ac$ is commutative
it admits non trivial geometries and these are
classified, up to homotopy of the operator $D$, by
integral quadratic forms $q=(q_{ij})_{i,j\in 1,\ldots
,N}$ which are invertible over $\Zb$. The \break geometry
associated to a quadratic form $q$ is defined as follows.
One lets $\Hc = \ \build \op_{i,j}^{} \Hc_{ij}$ where
$\Hc_{ij} = \Cb^{n_{ij}} \ , \ n_{ij} = \vert q_{ij}
\vert$. The left action of $\Ac$ in $\Hc$ is given, with
obvious notations by
$$
(a \, \xi)_{ij} = a_i \, \xi_{ij} \qquad \fl \, a \in \Ac
\ , \ \xi \in \Hc \, .
$$
The $\Zb /2$ grading $\g$ is given by
$$
(\g \, \xi)_{ij} = \g_{ij} \, \xi_{ij} \qquad \g_{ij} =
\Sign \, q_{ij}
$$
and the isometric antilinear involution $J$ is given by,
$$
(J \, \xi)_{ij} = \ov{\xi}_{ji} \, .
$$
(It makes sense since $q_{ij} = q_{ji}$.)

\smallskip

\noindent One has $K_0 (\Ac) = \Zb^n$ and the
intersection pairing is given by the quadratic form $q$.

\smallskip

\noindent A choice of $D$ is determined by a function $m$
on the edges of the graph $\G$ defined as follows. The
vertices of $\G$ are the $(i,j)$ with $q_{ij} \ne 0$, the
edges of $\G$ are the $(i,j)$, $(k,\ell)$ for which
$\g_{ij} \, \g_{k\ell} =-1$ and $i=k$ or $j=\ell$. The
function $m$ on $\G^{(1)}$ has to satisfy $m_{ij,k\ell} =
\ov{m}_{k\ell ,ij}$ and $m_{ij,k\ell} = \ov{m}_{ji,\ell
k}$. One can find $D$ so that the obtained geometry is
connected iff the above graph is connected.

\medskip

As a next simple example let $\Ac = C^{\ify} (M,M_k
(\Cb))$ be the algebra of $k\ts k$ matrix valued
functions on a smooth compact Spin manifold $M$. Let $g$
be a Riemannian metric on $M$ and $\Hc = L^2 (M,S \ot M_k
(\Cb))$ be the Hilbert space of $L^2$ sections of the
tensor product $S\ot M_k (\Cb)$ of the Spinor bundle $S$
by $M_k (\Cb)$. Then $\Ac$ acts on $\Hc$ by left
multiplication,
$$
(a \, \xi) (x) = a(x) \, \xi (x) \qquad \fl \, x\in M \, .
$$
The real structure $J$ is given by
$$
J= C \ot *
$$
where $C$ is the charge conjugation on spinors and $*$ is
the adjoint operation $T \ra T^*$ on matrices. This
operation transforms the left multiplication operators
$\xi \ra a \, \xi$ on matrices into right multiplication
operators $\xi \ra \xi \, a^*$ so that one checks axiom
($7'$),
$$
[a,b^0] = 0 \qquad \fl \, a,b \in \Ac \ , \ b^0 = J \,
b^* \, J^{-1} \, .
$$
As a first choice of $D$ one can take $D = {\part \!\!\!
/}_M \ot 1$, the tensor product of the Dirac operator 
${\part \!\!\! /}_M$ on $M$ by the identity on $M_k
(\Cb)$. One can then easily check that one obtains a non
commutative geometry in this way. This example is
relevant to illustrate two facts which hold in general in
the non commutative case. The first is that the group
$\Aut (\Ac)$ of $*$ automorphisms of the algebra $\Ac$,
which plays in general the role of the group $\Diff (M)$
of diffeomorphisms of the manifold $M$ and acts on the
space of representations $\pi$ by composition, has a
natural normal subgroup
$$
\Int M \sbs \Aut M
$$
where $\Int M$ is the group of inner automorphisms, i.e.
automorphisms of the form,
$$
\a_u (x) = u \, x \, u^* \qquad \fl \, x \in \Ac \, ,
$$
where $u$ is an arbitrary element of the unitary group
$$
\Uc = \{ u \in \Ac \ ; \ u \, u^* = u^* \, u =1\} \, .
$$
Moreover the action of this group of {\it internal
diffeomorphisms} on our geometries can be expressed in a
simple manner. Indeed the replacement of the
representation $\pi$ by $\pi \circ \a_u^{-1}$ is
equivalent to the replacement of the operator $D$ by
$$
\wt D = D + A + J \, A \, J^{-1}
$$
where $A = u [D,u^*]$.

\smallskip

\noindent The desired unitary equivalence is given by the
operator,
$$
U = u \, J \, u \, J^{-1} \, .
$$
(One checks that $U \, D \, U^* = \wt D$ and that $U \,
\a_u^{-1} (x) \, U^* =x$ for any $x \in \Ac$.)

\smallskip

\noindent The above perturbation of the metric is a
special case of {\it internal perturbation} given by,
$$
\wt D = D + A + J \, A \, J^{-1}
$$
where $A=A^*$ is now an arbitrary operator of the form
$$
A = \Si \, a_i \, [D,b_i] \qquad a_i ,b_i \in \Ac \, .
$$
In the commutative case such perturbations of the metric
all vanish because $A=A^*$ implies $A + J \, A \, J^{-1}
=0$. But in the non commutative case they do not. Thus in
our example with $k>1$ one gets that,
$$
\Int M = C^{\ify} (M, SU(k))
$$
is the group of local gauge transformations (internal
symmetries) for an $SU(k)$ gauge theory on $M$. The
internal perturbations of the metric are parametrized by
$SU(k)$ gauge potentials, i.e. by the non tracial part of
$A = \Si \, a_i \, [D,b_i]$, $a_i ,b_i \in \Ac$. One can
compute the effect of such internal perturbations to the
distance between two pure states $\vp ,\psi$ on $\Ac$. We
take $x,y \in M$ and let $\vp$ and $\psi$ correspond to
two unit vectors (rays) in $\Cb^k$ by the equality
$$
\vp (a) = \lgl a(x) \, \xi ,\xi \rgl \ , \ \psi (a) =
\lgl a(y) \, \eta ,\eta \rgl \qquad \fl \, a \in \Ac \, .
$$
The distance $d(\vp ,\psi)$ for the metric $\wt D$ is
defined as usual by
$$
d(\vp ,\psi) = \Sup \, \{ \vert \vp (a) - \psi (a)\vert \
; \ a\in \Ac \ , \ \Vert [\wt D ,a]\Vert \leq 1 \} \, .
$$
This distance depends heavily on the gauge connection
$A$. The latter defines an horizontal distribution $H$ on
the fibre bundle $P$ over $M$ whose fiber over each $x\in
M$ is the pure state space $P_{k-1} (\Cb)$ of $M_k (\Cb)$.
The metric $d$ turns out to be equal to the Carnot metric
([G]) on $P$ defined by the horizontal distribution $H$
and the Euclidean structure on $H$ given by the
Riemannian metric of $M$,
$$
d(\vp ,\psi) = \Inf \int_0^1 \Vert \pi_* (\dot{\g}
(t))\Vert \, dt \ , \ \g (0) = \vp \ , \ \g (1) =\psi
$$
where $\pi : P\ra M$ is the projection and $\g$ varies
through all {\it horizontal} paths $(\dot{\g} (t) \in H
\quad \fl \, t)$ which join $\vp$ to $\psi$. In
particular the finiteness of $d(\vp ,\psi)$ for $x=y$ is
governed by the holonomy of the connection at the point
$x$. The flat situation $A=0$ corresponds to the product
geometry of $M$ by the finite geometry where the algebra
is $M_N (\Cb)$ and $D=0$. For the latter since $D=0$ the
distance between any two $\vp \ne \psi$ is $+\ify$. Thus
for the product one gets that the connected components of
the metric topology are the flat sections of the bundle
$P$. Alternatively, if the holonomy at $x$ is $SU(k)$,
the metric on the fiber $P_x$ is finite.

\medskip

Our next example will be highly non commutative. It is a
special case of a general construction that works for any
isometry of a Riemannian Spin manifold, but we shall work
it out in the very specific case where this isometry is
the irrational rotation $R_{\t}$ of the circle $S^1$. This
will have the advantage of definitness but the whole
discussion is general. The algebra is the irrational
rotation smooth algebra,
$$
\Ac_{\t} = \{ \Si \, a_{nm} \, U^n \, V^m \ ; \
a=(a_{n,m}) \in \Sc (\Zb^2)\}
$$
where $\Sc (\Zb^2)$, the Schwartz space of $\Zb^2$, is the
space of sequences of rapid decay (i.e. $(1+\vert
n\vert +\vert m\vert)^k \, a_{n,m}$ is bounded for any $k
\geq 0$). The $*$ algebra structure is governed by the
presentation,
$$
U^* = U^{-1} \ , \ V^* = V^{-1} \ , \ V \, U = \lb \, U
\, V \ \hbox{with} \ \lb = \exp (2\pi i\t)
$$
with $\t \in \, ]0,1]$.

\smallskip

\noindent In order to specify the metric structure of our
non commutative geometry we shall need, as for usual
elliptic curves, a complex number $\tau$, $\Im \tau > 0$.
We can then describe the geometry as follows: the Hilbert
space $\Hc$ is given by the sum of 2-copies of $L^2
(\Ac_{\t} ,\tau_0)$ where $\tau_0$ is the canonical
normalized trace,
$$
\tau_0 (a) = a_{00} \qquad \fl \, a \in \Ac_{\t} \, .
$$
The Hilbert space $L^2 (\Ac_{\t} ,\tau_0)$ is just the
completion of $\Ac_{\t}$ for the inner product $\lgl a,b
\rgl = \tau_0 (b^* \, a) \quad \fl \, a,b \in \Ac_{\t}$.

\smallskip

\noindent The representation of $\Ac_{\t}$ in $\Hc$ is
given by left multiplication, i.e. as 2 copies of the
left regular representation,
$$
a \ra \left[ \matrix{ \lb (a) &0 \cr 0 &\lb(a) \cr}
\right] \ , \ \lb (a) \, b = a \, b \qquad \fl \, a,b \in
\Ac_{\t} \, .
$$
The operator $D$ depends explicitly on the choice of
$\tau$ and is
$$
D = \left[ \matrix{ 0 &\d_1 + \tau \, \d_2 \cr -\d_1
-\ov{\tau} \, \d_2 &0 \cr}\right]
$$
where the $\d_j$ are the following derivations of
$\Ac_{\t}$,
$$
\d_1 (U) = 2\pi i \, U \ , \ \d_1 (V) = 0 \ ; \ \d_2 (U) =
0 \ , \ \d_2 (V) = 2\pi i \, V \, .
$$
The $\Zb /2$ grading $\g$ is just $\g = \left[ \matrix{ 1
&0 \cr 0 &-1 \cr}\right]$.

\smallskip

\noindent To specify our geometry it remains to give the
antilinear isometry. We let $J=\left[ \matrix{ 0 &J_0 \cr
-J_0 &0 \cr}\right]$ where $J_0$ is Tomita's involution
([Ta]),
$$
J_0 \, a = a^* \qquad \fl \, a \in L^2 (\Ac_{\t} ,\tau_0)
\, .
$$
Note that $J_0 \, \lb (a^*) \, J_0^{-1} = \rho (a)$ is
the {\bf right} multiplication by $a\in \Ac_{\t}$ so that
our axiom ($7'$) is easy to check. The dimension $n$ is
equal to 2 so that $J \, D = D \, J$, $J^2 = -1$ and $J
\, \g = -\g \, J$ as expected.

\smallskip

\noindent The other axioms are easy to check but axiom
($4'$) is not so trivial because it requires to exhibit a
Hochschild 2-cycle, $c \in Z_2 (\Ac_{\t} ,\Ac_{\t})$ such
that $\pi (c) =\g$. It turns out to be,
$$
c = (2i\pi)^{-2} \, (\tau - \ov{\tau})^{-1} \, (V^{-1} \,
U^{-1} \ot U \ot V - U^{-1} \, V^{-1} \ot V \ot U) \, .
$$
(Note that this is not the same as the antisymmetrization
of $V^{-1} \, U^{-1} \ot U \ot V$ because of the phase
factor in $V^{-1} \, U^{-1} \ne U^{-1} \, V^{-1}$.) One
checks that $b \, c =0$ and $\pi (c) =\g$.

\smallskip

\noindent Note also that the area, $\mint \, ds^2 = \mint
\, D^{-2}$ of the non commutative torus depends on $\tau$
in the same way as $c$, the reason is that it has
homological significance,
$$
{\textstyle \mint} \, ds^2 = \lgl \vp , c \rgl
$$
where the cyclic cocycle $\vp$ is rigidly fixed by its
integrality as the chern character of our $K$-cycle.

\smallskip

\noindent It is also worthwile to check in some detail
the Poincar\'e duality. The point is that while at a
superficial level our geometry $\Tb_{\t}^2$ looks like an
ordinary torus, this impression quickly fades away if we
realize that for $\t \notin \Qb$ the algebra $\Ac_{\t}$
contains many non trivial idempotents ([Ri] [Po]) and
smooth real elements $a=a^*$ of $\Ac_{\t}$ can have
Cantor spectrum (thus from the point of view of $a$ the
space $\Tb_{\t}^2$ looks as totally disconnected). In
fact at first it might seem that Poincar\'e duality might
fail since there is no element $x\in K_0 (\Ac_{\t})$,
$x\ne 0$ of virtual dimension $0$, $\tau_0 (x) =0$, the
characteristic property of the Bott element. Indeed, when
$\t \notin \Qb$, the $K_0$ group is $\Zb^2$ ([Pi-V]) and
the trace $\tau_0 (n \, e_0 + m \, e_1) = n+m \, \t$
never vanishes unless $n=m=0$. The above specifies
uniquely the basis $e_0 = [1]$ and $e_1$, $\tau_0 (e_1) =
\t$ of $K_0$ and we take the basis $U,V$ for $K_1$. In
this basis the intersection form of axiom (6) is given by
the symplectic matrix,
$$
\left[ \matrix{
0 &-1 &0 &0 \cr
1 &0 &0 &0 \cr
0 &0 &0 &-1 \cr
0 &0 &1 &0 \cr
} \right] \, .
$$
It follows that Poincar\'e duality is satisfied and that
the Bott element $\b$ makes sense, as an element of $K_0
(\Ac_{\t}^0 \ot \Ac_{\t})$. It is given by
$$
\b = e_0^0 \ot e_1 - e_1^0 \ot e_0 + u^0 \ot v - v^0 \ot
u
$$
where we denote by $\ot$ the external cup product in $K$
theory. One can check that with $\mu$ the $K$-homology
class on $\Ac_{\t} \ot \Ac_{\t}^0$ given by our $K$-cycle
one has the Poincar\'e duality equation ([CS])
$$
\mu \ot_{\Ac_{\t}^0} \, \b = \id_{\Ac_{\t}} \qquad \b
\ot_{\Ac_{\t}} \, \mu = \id_{\Ac_{\t}^0} \, .
$$
Finally the unique trace on $\Ac_{\t}^0 \, \ot \,
\Ac_{\t}$ is $\tau_0^0 \, \ot \, \tau_0$ and it does vanish
on $\b$, $\lgl \tau_0^0 \, \ot \, \tau_0 ,\b \rgl =0$ which
is the expected property of the Bott element.

\smallskip

\noindent In the general theory both $\b \in K_*
(\Ac_{\t}^0 \, \ot \, \Ac_{\t})$ and $\mu \in K^* (\Ac_{\t}
\, \ot \, \Ac_{\t}^0)$ make sense and satisfy the above
equation. Moreover the local index theorem of [CM] allows
to compute the pairing $\lgl \mu ,\b \rgl$ by a local
formula, i.e. a formula invoking $\mint$ of simple
algebraic expressions. This yields the Gauss Bonnet
theorem in our context, due to the equality,
$$
\lgl \mu ,\b \rgl = \Rank K_0 - \Rank K_1
$$
(where the rank is the rank of the abelian group, i.e.
$\dim_{\Qb} (K_j \ot \Qb )$). Using the automorphism $\s$
of $\Ac_{\t}$ associated to a unimodular integral matrix
$\left[ \matrix{ a &b \cr c &d \cr} \right] \in
SL(2,\Zb)$ by,
$$
\s (U) = U^a \, V^b \ , \ \s (V) = U^c \, V^d
$$
it is not difficult to check that, up to an irrelevant
scale factor, the non commutative geometry $\Tb_{\t
,\tau}^2$ only depends upon the value of $\tau$ in $\Cb^+
/ PSL(2,\Zb)$. The new phenomenon which occurs in the non
commutative case is the {\it Morita equivalence} of
geometries which will correspond here to the action of
the modular group $PSL (2,\Zb)$ on the module $\t$ rather
than $\tau$. To change $\t$ in this way one uses a finite
projective right module $\Ec$ over the algebra $\Ac$ and
then one replaces $\Ac$ by the algebra $\Bc = \End_{\Ac}
(\Ec)$. To be more specific let us change $\t$ in
$-{1\over \t}$ as follows. We take for $\Ec$ the
following right $\Ac$-module ([Co]). As a vector space
$\Ec$ is the Schwartz space $\Sc (\Rb)$ of functions on
$\Rb$. The right action of $\Ac_{\t}$ on $\Sc$ is
specified by the rules,
$$
(\xi \, U)(s) = \xi (s+\t) \qquad \fl \, \xi \in \Sc (\Rb)
\ , \ \fl \, s \in \Rb
$$
$$
(\xi \, V)(s) = e^{2\pi is} \, \xi (s) \qquad \fl \, \xi
\in \Sc (\Rb) \ , \ \fl \, s \in \Rb \, .
$$
One checks directly that the algebra $\Bc = \End_{\Ac_{\t}}
(\Sc)$ is isomorphic to $\Ac_{-1/\t}$ with generators
given by the translation by 1 and the multiplication by
the function $s \ra \exp \left( {2\pi is \over \t}
\right)$. This $\Ac_{\t}$- module $\Sc$ is naturally a
hermitian module ([Co]) and we can now apply the
following general operation of Morita equivalence in non
commutative geometry.

\smallskip

\noindent Let $(\Ac ,\Hc ,D)$ be a given geometry and
$\Bc=\End_{\Ac} (\Ec)$ a Morita equivalent algebra. Then a
non commutative geometry, $(\Bc ,\wt{\Hc} ,\wt D)$ is
uniquely specified by a {\it hermitian connection} ([Co])
$\nb$ on $\Ec$. Such a connection is a linear map,
$$
\nb : \Ec \ra \Ec \ot_{\Ac} \Om_D^1
$$
where $\Om_D^1$ is the $\Ac$-bimodule,
$$
\Om_D^1 = \{ \Si \, a_i \, [D,b_i] \ ; \ a_i ,b_i \in \Ac
\} \sbs \Lc (\Hc)
$$
and $\nb$ should satisfy the Leibnitz rule and the
compatibility with the hermitian structure,

\smallskip

\noindent 1) $\nb (\xi \, a) = (\nb \, \xi) \, a + \xi
\ot [D,a] \qquad \fl \, a\in \Ac \ , \ \xi \in \Ec$

\smallskip

\noindent 2) $(\xi ,\nb \, \eta) - (\nb \, \xi ,\eta) =
d(\xi ,\eta) \qquad \fl \, \xi ,\eta \in \Ec$

\smallskip

\noindent where $d \, a = [D,a]$ by definition.

\smallskip

\noindent To construct $\wt{\Hc}$ and $\wt D$ one
proceeds as follows.

\smallskip

\noindent One lets $\wt{\Hc} = \Ec \ot_{\Ac} \Hc
\ot_{\Ac} \ov{\Ec}$ where $\ov{\Ec}$ is the conjugate
module, with elements $\ov{\xi}$, $\xi \in \Ec$ and the
module structure,
$$
a \, \ov{\xi} = (\xi \, a^*)^- \qquad \fl \, \xi \in \Ec
\ , \ a\in \Ac \, .
$$
Then $\wt{\Hc}$ has a natural Hilbert space structure
(cf. [Co]) obtained using the hermitian structure
of $\Ec$.

\smallskip

\noindent The operator $\wt D$ is given by the formula,
$$
\wt D (\xi \ot \eta \ot \ov{\z}) = (\nb \, \xi) \, \eta
\ot \ov{\z} + \xi \ot D \, \eta \ot \ov{\z} + \xi \ot
\eta (\ov{\nb \, \z})
$$
where we take advantage of the action of $\Om_D^1$ in
$\Hc$ in order to make sense of $(\nb \, \xi) \, \eta$
for instance (cf. [Co]). The above formula is compatible
with tensor products over $\Ac$. Similarly the real
structure $\wt J$ is given by (with obvious notations),
$$
\wt J \, (\xi \ot \eta \ot \ov{\z}) = \z \ot J \, \eta
\ot \ov{\xi} \, .
$$
One then checks that all our axioms are fulfilled by the
triple $(\Bc ,\wt{\Hc} ,\wt D)$ and $\wt J$ where the
action of $\Bc$ is given by
$$
b \, (\xi \ot \eta \ot \ov{\z}) = (b \, \xi) \ot \eta
\ot \ov{\z} \, .
$$
In our specific example of the right module $\Sc$ on
$\Ac_{\t}$ the $\Ac_{\t}$ bimodule $\Om_D^1$ is easily
computed and (cf. [Co]) as a bimodule it is the sum of two
copies of $\Ac_{\t}$ which we write by identifying $\om
\in \Om_D^1$ with a $2\ts 2$ matrix of the form,
$$
\om = \left[ \matrix{
0 &\lb (a) \cr \lb (b) &0 \cr
} \right]
$$
and writing $a=(\d_1 + \tau \, \d_2) (x)$, $b=(-\d_1
,-\ov{\tau} \, \d_2) (x)$ for any $x\in \Ac$ and $\om =
dx$. A connection $\nb$ is uniquely specified by the two
covariant derivatives $\nb_j$ which correspond to $\d_j$.
Thus a specific choice of connection on $\Sc$ is given by 
$$
\eqalign{
(\nb_1 \, \xi) (s) = & \ {2\pi i \over \t} \, s \, \xi
(s) \cr
(\nb_2 \, \xi) (s) = & \ \xi' (s) \, . \cr
}
$$
We leave as an exercise to compute the corresponding
value of the module $\tau'$ for the Morita equivalent
geometry on $\Tb_{-1/\t}^2$. The connection $\nb$ used in
this example has constant curvature (cf. [C-R]) but there
is a lot of freedom in the choice of the connection.
Indeed the space of connections is naturally an affine
space over the vector space $\Hom_{\Ac} (\Ec ,\Ec
\ot_{\Ac} \Om_D^1)$ (with the selfadjointness condition).

\smallskip

\noindent All this discussion applies in particular when
$\Ec =\Ac$ and yields a whole new class of geometries,
the {\it internal perturbation} which replace $D$ by
$D+A+J \, A \, J^{-1}$ where $A$ is an arbitrary
selfadjoint element of $\Om_D^1$.

\smallskip

\noindent These internal perturbations are trivial: $A+J
\, A \, J^{-1} =0$ when the geometry is commutative
(theorem 1). They are highly non trivial for
$\Tb_{\t}^2$, $\t \notin \Qb$.

\smallskip

\noindent When $\t \notin \Qb$ the von Neumann algebra
weak closure of $\Ac_{\t}$ in $\Hc$ is the hyperfinite
factor of type $\Tb_1$ and the antilinear isometry $J_0$
is Tomita's involution. In the general theory it is
always true that the von Neumann algebra $\Ac''$ weak
closure of $\Ac$ in $\Hc$ is finite and hyperfinite. This
follows from axiom 5, the general properties of the
Dixmier trace and the results of [Co]. In particular the
von Neumann algebras involved in our geometries are
completely classified up to isomorphism ([Co]).

\bigskip

\noindent {\it The non commutative geometry of the
standard model.}

\smallskip

\noindent We shall now describe a simple {\it finite}
geometry of dimension equal to 0, $(\Ac_F ,\Hc_F ,$ $D_F)$
($F$ for finite), whose product by ordinary Euclidean
4-dimensional geometry (or more generally by a
4-dimensional Spin manifold) will give the standard model
(SM) in the following way:

\smallskip

\item{1)} The Hilbert space $\Hc$ will describe the (one
particle) Fermionic sector of the SM.

\smallskip

\item{2)} The inner fluctuations of the metric $D\ra D +
A + J \, A \, J^{-1}$ give exactly the bosonic sector of
SM with the correct quantum numbers and hypercharges for
the coupling with the fermions: $\lgl D\psi ,\psi \rgl$.

\item{3)} The {\it spectral action} Trace $(\vp (D^{-1}))
+ \lgl D\psi , \psi \rgl$ restricted to the inner
fluctuations of the metric, gives the SM Lagrangian.

\smallskip

\noindent We postpone the proof of (3) and the analysis of
its relation to gravity to our collaboration with A.C.
[CC].

\smallskip

\noindent In this paper we shall describe the geometry
$(\Ac_F ,\Hc_F ,D_F)$ and check (2) in some detail.

\smallskip

\noindent We let $\Hc_F$ be the Hilbert space with basis
the list of elementary Fermions. Thus each generation of
Fermions contributes by a space of dimension $15+15$
where $15=12+3$. The 12 corresponds to the quarks and
$12=4\ts 3$ where the 4 is given by the table
$$
\matrix{
u_R &u_L \cr
d_R &d_L \cr
}
$$
of up and down particles of left or right chirality,
while the 3 is given by the color index. The 3 in
$15=12+3$ corresponds to the leptons and is given by the
table
$$
\matrix{
&\nu_L \cr
e_R &e_L \cr
} \, .
$$
The second 15 in $15+15$ corresponds to antiparticles and
is obtained by putting an $\ov f$ instead of an $f$ for
any $f$ in the above basis. This gives us the antilinear
isometry $J=J_F$ in $\Hc_F$, by
$$
J_F (\Si \, \lb_i \, f_i + \Si \, \mu_j \, \ov{f}_j) =
\Si \, \ov{\mu}_j \, f_j + \Si \, \ov{\lb}_i \, \ov{f}_i
$$
for any $\lb_i ,\mu_j \in \Cb$.

\smallskip

\noindent We let $\Ac_F = \Cb \op \Hb \op M_3 (\Cb)$ be
the direct sum the real involutive algebras $\Cb$ of
complex numbers, $\Hb$ of quaternions, and $M_3 (\Cb)$ of
$3\ts 3$ matrices. (Recall that quaternions $q$ can be
represented as $2 \ts 2$ matrices of the form $\left[ {\a
\over -\b} \ {\b \over \a} \right]$ where $\a ,\b \in
\Cb$.)

\smallskip

\noindent Let us give the action of $\Ac$ in $\Hc_F$. We
let, for $a=(\lb ,q,m) \in \Ac$,
$$
\matrix{
a \, u_R = \lb \, u_R & \qquad a \, u_L = \a \, u_L -
\ov{\b} \, d_L \cr
\cr
a \, d_R = \ov{\lb} \, d_R &\qquad a \, d_L = \b \, u_L +
\ov{\a} \, d_L \, . \cr
}
$$
(Independently of the color index), while for leptons the
formula is the same but there is no $u_R$.

\smallskip

\noindent This fixes completely the action of $\Ac$ on
particles. The action on antiparticles $\ov f$ is given
by: (for $a=(\lb ,q,m)$ as above)
$$
\matrix{
a \, \ov f = \lb \, \ov f &\hbox{if} \ f \ \hbox{is a
lepton} \cr
\cr
a \, \ov f = m \, \ov f &\hbox{if} \ f \ \hbox{is a
quark} \, . \cr
}
$$
Here the $3\ts 3$ matrix is acting in the obvious way on
the color index.

\smallskip

\noindent For the operator $D_F$ we take $D_F = \left[
\matrix{ Y &0 \cr 0 &\ov Y \cr} \right]$ where $Y$ is the
Yukawa coupling matrix, which has the dimension of a
mass, and is of the form:
$$
Y = Y_q \ot 1_3 \op Y_f
$$
with
$$
Y_q = \left[ \matrix{
0 &0 &M_u &0 \cr
0 &0 &0 &M_d \cr
M_u^* &0 &0 &0 \cr
0 &M_d^* &0 &0 \cr
} \right] \ , \ Y_f = \left[ \matrix{
0 &0 &M_e \cr
0 &0 &0 \cr
M_e^* &0 &0 \cr
} \right] \, .
$$
For one generation $M_u ,M_d$ and $M_e$ would just be
scalars but for 3 generations they are matrices which
encode both the masses of the Fermions and their mixing
properties.

\smallskip

\noindent Let us now check our axioms for this
0-dimensional geometry $(\Ac_F ,\Hc_F ,D_F)$. We begin by
checking $7')$. One has $J_F^2 =1$, $J_F \, D_F = D_F \,
J_F$ and it is clear that $J_F$ commutes with the natural
$\Zb /2$ grading given by chirality:
$$
\g (f_R) = f_R \ , \ \g (f_L) = -f_L \qquad (f \
\hbox{particle or antiparticle}) \, .
$$
But the important property is that $[a,b^0] = 0$ for any
$a,b \in \Ac$ (cf. $(7')$), where $b^0 = J \, b^* \,
J^{-1} \quad \fl \, b\in \Ac$. To check it, it is enough
to see what happens on one generation of particles. For
quarks the right action $b^0$ of $b\in \Ac \quad b = (\lb
,q,m)$ is given by $m^t$ acting on the color index which
obviously commutes with the left action of $\Ac$. For
leptons the right action is by the scalar $\lb$ which
also commutes with the left action of $\Ac$.

\smallskip

\noindent We have thus checked $(7')$. Since $n=0$ and
$\Hc_F$ is finite dimensional the axiom (1) is obvious.
To check $(2')$ we need to show that $[[D,a],b^0] = 0
\quad \fl \, a,b \in \Ac$.

\smallskip

\noindent Once again it is enough to check it for one
generation of particles, it is clear for leptons and it is
true for quarks exactly because the color is unbroken so
that both $a$ and $D$ exactly commute with the right
action of $\Ac$.

\smallskip

\noindent Being in finite dimension the axiom 3 is
obvious.

\smallskip

\noindent To check $(4')$ one verifies that $\g = \ve \,
J \, \ve \, J$ where $\ve$ is the following element of
$\Ac$, $\ve = (1,-1,1)$. Thus one has $c=\ve \ot \ve^0
\in \Ac \ot \Ac^0$.

\smallskip

\noindent Note that our algebra $\Ac$ is real so that (5)
has to be stated for the complex algebra generated by
$\Ac$ in $\Hc_F$. It is then clear. Finally when we
compute the intersection form on $K_0 (\Ac) = \Zb \op \Zb
\op \Zb$ we find, with $N$ the number of generators, the
$3\ts 3$ matrix,
$$
Q = 2N \ \left[ \matrix{
-1 &1 &-1 \cr
1 &0 &1 \cr
-1 &1 &0 \cr
} \right] \, .
$$
The above matrix is invertible with inverse given by
$$
Q^{-1} = (2N)^{-1} \ \left[ \matrix{
1 &1 &-1 \cr
1 &1 &0 \cr
-1 &0 &1 \cr
} \right]
$$
so that (6) only holds rationally.

\smallskip

\noindent We now consider a 4-dimensional smooth compact
Riemannian manifold $M$ with a fixed spin structure and
consider its product with the above finite geometry. One
can prove that our notion of geometry is stable by
products. When one of the two geometries is even (i.e. it
possesses a $\Zb /2$ grading $\g_1$), the product
geometry is given by the rules,
$$
\Ac = \Ac_1 \ot \Ac_2 \ , \ \Hc = \Hc_1 \ot \Hc_2 \ , \ D
= D_1 \ot 1 + \g_1 \ot D_2 \, .
$$
To check axiom $(4')$ for instance one uses the shuffle
product in Hochschild homo\-logy (cf. Loday).

\smallskip

\noindent For the product of the manifold $M$ by the
finite geometry $F$ we thus have $\Ac = C^{\ify} (M) \ot
\Ac_F = C^{\ify} (M,\Ac_F)$, $\Hc = L^2 (M,S) \ot \Hc_F =
L^2 (M,S \ot \Hc_F)$ and $D = {\part \!\!\! /}_M \ot 1 +
\g_5 \ot D_F$ where ${\part \!\!\! /}_M$ is the Dirac
operator on $M$.

\smallskip

\noindent Let us check that the inner fluctuations of the
metric give us the gauge bosons of the standard model
with their correct quantum numbers. We first have to
compute $A = \Si \, a_i [D,a'_i] \quad a_i ,a'_i \in
\Ac$. Since $D = {\part \!\!\! /}_M \ot 1 + \g_5 \ot D_F$
decomposes as a sum of two terms, so does $A$ and we
first consider the discrete part $A^{(0,1)}$ coming from
commutators with $\g_5 \ot D_F$. Let $x\in M$ and let
$a_i (x) = (\lb_i , q_i ,m_i)$, $a'_i (x) = (\lb'_i ,q'_i
,m'_i)$, the computation of $\Si \, a_i [\g_5 \ot D_F
,a'_i]$ at $x$ gives $\g_5$ tensored by the following
matrices,
$$
\left[ \matrix{
0 &X \cr
X' &0 \cr
} \right] \ , \ X = \left[ \matrix{
M_u \, \vp_1 &M_u \, \vp_2 \cr
-M_d \, \ov{\vp}_2 &M_d \, \ov{\vp}_1 \cr
} \right] \ , \ X' = \left[ \matrix{
M_u^* \, \vp'_1 &M_d^* \, \vp'_2 \cr
-M_u^* \, \ov{\vp}'_2 &M_d^* \, \ov{\vp}'_1 \cr
} \right]
$$
for the quark part, with $\vp_1 = \Si \, \lb_i (\a'_i -
\lb'_i)$, $\vp_2 = \Si \, \lb_i \, \b'_i$
$$
\vp'_1 = \Si \, \a_i (\lb'_i - \a'_i) + \b_i \,
\ov{\b}'_i \ , \ \vp'_2 = \Si \, (-\a_i \, \b'_i + \b_i
(\ov{\lb}'_i - \ov{\a}'_i )) \, .
$$
For the lepton part one gets the $3\ts 3$ matrix,
$$
\left[ \matrix{
0 &-M_d \, \ov{\vp}_2 & M_d \, \ov{\vp}_1 \cr
M_d^* \, \vp'_2 &0 &0 \cr
M_d^* \, \ov{\vp}'_1 &0 &0 \cr
} \right]
$$
where $\vp_1$, $\vp_2$, $\vp'_1$ and $\vp'_2$ are as
above.

\smallskip

\noindent Let $q=\vp_1 + \vp_2 \, j$, $q' = \vp'_1 +
\vp'_2 \, j$ where $j$ is the quaternion $\left[ \matrix{
0 &1 \cr -1 &0 \cr} \right]$. The selfadjointness
condition $A=A^*$ is equivalent to $q' = q^*$ and we see
that the discrete part $A^{(0,1)}$ is exactly given by a
quaternion valued function, $q(x) \in \Hb$ on $M$. This
pair of complex fields is the Higgs doublet and one
checks that it has the right quantum numbers. The
antiparticle sector does not contribute to $A^{(0,1)}$
because the left action of $\Ac_F$ on this sector exactly
commutes with $D_F$. 

\smallskip

\noindent Let us now determine the other part $A^{(1,0)}$
of $A$, i.e.
$$
A^{(1,0)} = \Si \, a_i [({\part \!\!\! /}_M \ot 1), a'_i]
\, .
$$
With obvious notations, $a_i = (\lb_i ,q_i ,m_i)$, $a'_i
= (\lb'_i ,q'_i ,m'_i)$ we obtain,

\smallskip

A $U(1)$ gauge field $\L = \Si \, \lb_i \, d \, \lb'_i$

\smallskip

A $SU(2)$ gauge field $Q=\Si \, q_i \, d \, q'_i$

\smallskip

A $U(3)$ gauge field $V = \Si \, m_i \, d \, m'_i$.

\smallskip

\noindent The computation of $A + J \, A \, J^{-1}$ gives
the following matrices on quarks and leptons, where we
omit the symbol of Clifford multiplication,
$$
\left[ \matrix{
\L + V &0 &0 &0 \cr
0 &-\L + V &0 &0 \cr
0 &0 &Q_{11} + V &Q_{12} \cr
0 &0 &Q_{21} &Q_{22} + V \cr
} \right]
$$
$$
\left[ \matrix{
-2\L &0 &0 \cr
0 &Q_{11} - \L &Q_{12} \cr
0 &Q_{21} &Q_{22} -\L \cr
} \right]
$$
where the matrix for quarks is a $4\ts 4$ matrix of $3\ts
3$ matrices because $V$ is a $3\ts 3$ matrix, (ignoring
the flavor index). Since we are only interested in the
fluctuation of the metric we shall write the total $15
\ts 15$ matrix as the sum of a traceless matrix plus a
scalar multiple of the identity matrix. Since the latter
does not affect the metric we shall ignore it. This
amounts to remove a $U(1)$ by imposing the condition that
the full matrix is traceless, i.e. that 4 trace $V - 4\L
=0$, i.e. trace $V=\L$. Thus we get $V=V' + {1\over 3} \,
\L$ with trace $V'=0$ so that $V'$ is an $SU(3)$ gauge
potential. The $U(1)$ field $\L$ is the generator of
hypercharge and we obtain the following matrices for the
inner fluctuation $A + J \, A \, J^{-1}$ of the metric
(vector part),
$$
\matrix{
u_R \cr d_R \cr u_L \cr d_L \cr
} \ \left[ \matrix{
{4\over 3} \, \L + V' &0 &0 &0 \cr
0 &-{2\over 3} \, \L + V' &0 &0 \cr
0 &0 &Q_{11} + {1\over 3} \, \L + V' &Q_{12} \cr
0 &0 &Q_{21} &Q_{22} + {1\over 3} \, \L + V' \cr
} \right]
$$
$$
\matrix{
e_R \cr \nu_L \cr e_L \cr
} \ \left[ \matrix{
-2\L &0 &0 \cr
0 &Q_{11} - \L &Q_{12} \cr
0 &Q_{21} &Q_{22} -\L \cr
} \right] \, .
$$
Together with the above Higgs doublet which gives the
scalar part of $A+J \, A \, J^{-1}$ we thus obtain
exactly the gauge bosons of the standard model coupled
with the correct hypercharges $Y_L , Y_R$. They are such
that the electromagnetic charge $Q_{em}$, is determined
by $2 \, Q_{em} = Y_R$ for right handed particles and $2
\, Q_{em} = Y_L + 2 \, I_3$ where $I_3$ is the 3\up{rd}
generator of the weak isospin group $SU(2)$. For $Q_{em}$
one gets the same answer for the left and right
components of each particle and ${2\over 3}$, $-{1\over
3}$ for $u,d$ respectively and $0,-1$ for $\nu$ and $e$
respectively.

\smallskip

\noindent We showed in [C] that one obtains the full
Lagrangian of the standard model from the sum $\mint \,
\t^2 \, ds^4 + \lgl (D+A+J \, A \, J^{-1} ) \psi ,\psi
\rgl$ where $\t$ is the curvature of the connection $A$.
However this requires the definition of the curvature and
is still in the spirit of gauge theories. What the
present paper shows is that one should consider the
internal gauge symmetries as part of the diffeomorphism
group of the non commutative geometry, and the gauge
bosons as the internal fluctuations of the metric. It
follows then that the action functional should be of
purely gravitational nature. We state the principle of
spectral invariance, stronger than the invariance under
diffeomorphisms, which requires that the action
functional only depends on the spectral properties of $D
= ds^{-1}$ in $\Hc$. This is verified by the action,
$$
I = \hbox{Trace} \, (\vp (ds/ \ell_p)) + \lgl D \, \psi
,\psi \rgl
$$
for any nice function $\vp$ from $\Rb_+^*$ to $\Rb$.

\smallskip

\noindent We shall show in [CC] that this action gives
the SM Lagrangian coupled with gravity. It would seem at
first sight that the algebra $\Ac$ has disappeared from
the scene when one writes down the above action, the point
is that it is still there because it imposes the
constraints $[[D,a],b^0] =0 \quad \fl \, a,b \in \Ac$ and
$\Si \, a_i^0 [D,a_i^1] \ldots [D,a_i^4] = \g$ coming
from axioms 2 and 4.

\smallskip

\noindent It is important at this point to note that the
{\it integrality}, $n\in \Nb$ of the dimension of a non
commutative geometry appears to be essential to define
the Hochschild cycle $c\in Z_n$ and in turns the
chirality $\g$. This is very similar to the obstruction
which appears when one tries to apply dimensional
regularization to chiral gauge theories.

\smallskip

\noindent The relations (axioms 2 and 4) which relate the
algebra $\Ac$ with the infinitesimal length element $ds$
are very simple but it is clear that more involved
relations of a quartic type are necessary to cover the
hypoelliptic situation encountered in [CM]. Also the
algebra $\Ac = C^{\ify} (M) \ot \Ac_F$ used in the
description of the standard model is only slightly non
commutative. This commutativity should fade away when one
gets to energies $\L \sim m_p$ so that the inner
automorphisms $\Int \Ac$ exhaust more and more
automorphisms of $\Ac$. It is tempting to follow the
approximation scheme given in [GKP] but the main difficulty
is that the above non linear constraints between the
algebra $\Ac$ and the operator $D$ do not have finite
dimensional realizations since the Hochschild homology
$H_4 (\Ac ,\Ac)$ vanishes for any finite dimensional
algebra $\Ac$.

\smallskip

\noindent As a final remark we look for an explanation of
the remaining non commutativity of the algebra $\Ac =
C^{\ify} (M) \ot \Ac_F$ from the theory of quantum groups
at roots of unity. The simple fact is that the Spin
covering of $SO(4)$, i.e. $\Spin (4)$ is not the maximal
covering in the world of quantum groups. Indeed, $\Spin
(4) = SU(2) \ts SU(2)$ and even the single group $SU(2)$
admits, in the sense of non commutative geometry (cf.
[M]) non trivial extensions of finite order, (Frobenius
at $\ify$)
$$
1 \ra H \ra SU(2)_q \ra SU(2) \ra 1
$$
where $q$ is any root of unity, $q^m =1$, of odd order.
The simplest instance is when $m=3$ so that $q=\exp
\left( {2\pi i \over 3}\right)$. The quantum group $H$
has a finite dimensional Hopf algebra which is closely
related to the algebra $\Ac_F$. The Spin representation
of $H$ defines, like any other representation of $H$, a
bimodule over the Hopf algebra of $H$. The structure of
this bimodule turns out to be very similar to the
structure of the bimodule $\Hc_F$ over $\Ac_F$ that we
described above. The details of the adaptation of these
ideas to $\Spin (4) = SU (2) \ts SU(2)$ still remain to be
elucidated.

\vfill\eject

\noindent {\bf Bibliography}

\medskip

\item{[CC]} A. Chamseddine and A. Connes; The
spectral action principle, to appear.

\item{[C]} A. Connes; Non commutative geometry and
reality, {\it Journal of Math. Physics} {\bf 36} n\up{0}11
(1995).

\item{[Co]} A. Connes; {\it Non commutative geometry},
Academic Press 1994.

\item{[CM]} A. Connes and H. Moscovici; The local index
formula in non commutative geometry, GAFA.

\item{[CR]} A. Connes and M. Rieffel; Yang Mills for non
commutative two tori, {\it Operator algebras and
mathematical physics}, (Iowa City, Iowa 1985), 237-266.

\item{[CS]} A. Connes and G. Skandalis; The longitudinal
index theorem for foliations, {\it Publ. Res. Inst. Math.
Sci. Kyoto} {\bf 20} (1984), 1139-1183.

\item{[G]} M. Gromov; Carnot-Caratheodory spaces seen
from within, Preprint \break IHES/M/94/6.

\item{[GKP]} H. Grosse, C. Klimcik and P. Presnajder; On
finite 4 dimensional quantum field theory in
non commutative geometry, CERN Preprint TH/96-51,
hep-th/9602115.

\item{[K]} D. Kastler; The Dirac operator and
gravitation, {\it Commun. Math. Phys.} {\bf 166} (1995),
633-643.

\item{[KW]} W. Kalau and M. Walze; Gravity, non
commutative geometry and the \break Wodzicki residue, {\it
J. of Geom. and Phys.} {\bf 16} (1995), 327-344.

\item{[L]} J.L. Loday; {\it Cyclic homology}, Springer
Berlin 1992.

\item{[LM]} B. Lawson and M.L. Michelson; {\it Spin
Geometry}, Princeton 1989.

\item{[M]} Y. Manin; Quantum groups and non commutative
geometry, {\it Centre Recherche Math. Univ.
Montr\'eal} (1988).

\item{[PV]} M. Pimsner and D. Voiculescu; Exact sequences
for $K$ groups and Ext groups of certain crossed product
$C^*$ algebra, {\it J. Operator Theory} {\bf 4} (1980),
93-118.

\item{[S]} D. Sullivan; Geometric periodicity and the
invariants of manifolds, {\it Lecture Notes in Math.}
{\bf 197}, Springer (1971).

\item{[Ta]} M. Takesaki; Tomita's theory of modular
Hilbert algebras and its applications, {\it Lecture Notes
in Math.} {\bf 128}, Springer (1970).

\bye